\documentclass[aps,prl,twocolumn,groupedaddress,showpacs]{revtex4}

\usepackage{graphicx}
\begin{document}

\title{Glauber dynamics in a single-chain magnet: From theory to real systems}

\author{Claude Coulon$^{1}$, Rodolphe Cl\'erac$^1$, Lollita Lecren$^1$,
Wolfgang Wernsdorfer$^2$, Hitoshi Miyasaka$^{3,4}$}

\affiliation{
$^1$Centre de Recherche Paul Pascal, CNRS UPR 8641, 115 Avenue Dr. A.
Schweitzer, 33600 Pessac, France\\
$^2$Laboratoire Louis N\'eel, associ\'e \`a l'UJF, CNRS, BP 166,
38042 Grenoble Cedex 9, France\\
$^3$Department of Chemistry, Graduate School of Science, Tokyo
Metropolitan University, Minami-Ohsawa 1-1, Hachioji, Tokyo 192-0397\\
$^4$Structural Ordering and Physical Properties (PRESTO),
Science and Technology Agency (JST), \\ 
4-1-8 Honcho Kawaguchi, Saitama 332-0012, Japan 
}

\date{Submitted to PRL 10 May 2003. Submitted to PRB 12 December 2003; published 15 April 2004.}

\begin{abstract}
The Glauber dynamics is studied in a single-chain magnet (SCM).
As predicted a single relaxation mode of the magnetization is found.
Above 2.7 K, the thermally activated relaxation time is mainly governed
by the effect of magnetic correlations and the
energy barrier experienced by each magnetic unit.
This result is in perfect agreement with independent
thermodynamical measurements. Below 2.7 K, a crossover towards
a relaxation regime is observed
that is interpreted as the manifestation of finite-size effects.
The temperature
dependences of the relaxation time and of the magnetic susceptibility
reveal the importance of the boundary conditions.
\end{abstract}


\maketitle

The design of new slow-relaxing magnetic
nanosystems is a very challenging goal
for both applications (as information storage)
and fundamental research. A well-known example
of such systems is the single-molecule magnet (SMM)
that shows slow reversal of the magnetization
due to the combined effect of a high spin
ground state and uniaxial anisotropy
producing an energy barrier between
spin-up and spin-down states~\cite{Gatteschi03}.
When a magnetic field is initially applied
to magnetize this system and then removed,
the magnetization decays with a
material-inherent relaxation time depending
on the temperature. The corresponding relaxation
time, $\tau$, follows an Arrhenius law
at high temperatures and the activation
energy is equal to the barrier height,
being roughly $|D|S^2$, where $D$ is the
negative uniaxial anisotropy constant and $S$ is
the spin ground state of the molecule.
At lower temperatures, $\tau$ may saturate
when quantum tunneling through the barrier
becomes relevant~\cite{Barra01}.

Another research route of metastable magnetism
has recently been explored with the synthesis
of single-chain magnets (SCMs)
~\cite{Caneschi01,Clerac02,Lescouezec03}.
In these materials, the slow relaxation of magnetization
is not solely the consequence of the uniaxial
anisotropy seen by each spin on the chain
but depends also on magnetic correlations.
The effect of the short-range order becomes
more and more important when the temperature
is reduced until a critical point
is reached at T = 0 K for 1D systems.
In fact, the relaxation time is found to be
exponentially enhanced at low temperatures
in agreement with the pioneer work of
R. J. Glauber devoted to the dynamics of
the 1D Ising model~\cite{Glauber63}.
Although it seems that there is a reasonable agreement
between the experimental data and the Glauber's
theory~\cite{Caneschi01,Clerac02,Lescouezec03},
we show in this communication that several other arguments should be
considered to fill the gap between the theory and the experimental results.
Firstly, it should be mentioned that
the experimental systems are not strictly Ising-like.
In the simplest case,
they are rather described by an anisotropic Heisenberg model:
\begin{equation}
       \mathcal{H} = -2 J \sum_{i}\vec{S}_i\vec{S}_{i+1} +
                      D \sum_{i} S_{i,z}^2
\label{Heisenberg}
\end{equation}
where $J$ is the ferromagnetic exchange constant
between the spin units and $D$ is the single-ion
anisotropy. Secondly, the relaxation
time of each magnetic unit, introduced
phenomenologically in Glauber's study,
is {\it a priori} temperature dependent~\cite{Susuki68}
and this argument should also be considered.
This question is particularly important
if each magnetic unit is by itself a slowly relaxing object.
Finally, a real material is never perfect and, as the magnetic
correlation length becomes exponentially large at low temperature, even
a very small number of defects should deeply affect the magnetic behavior
of these one-dimensional systems~\cite{note60}.

\begin{figure}
\begin{center}
\includegraphics[width=.4\textwidth]{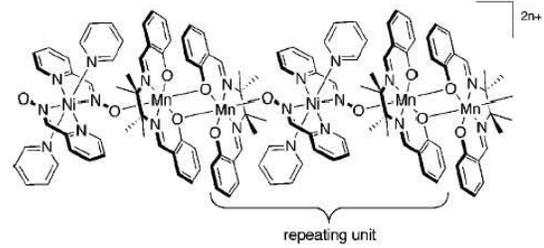}
\caption{Schematic view of the chain structure
showing the magnetic units (Mn-Ni-Mn trimers) in
[Mn$_2$(saltmen)$_2$Ni(pao)$_2$(py)$_2$](ClO$_4$)$_2$ [4].}
\label{schema}
\end{center}
\end{figure}

In this letter, we analyze the magnetic relaxation
of a 1D system,
[Mn$_2$(saltmen)$_2$Ni(pao)$_2$(py)$_2$](ClO$_4$)$_2$
(saltmen$^{2-}$ = N,N'-(1,1,2,2-tetramethylethylene)
bis(salicylideneiminate); pao$^{-}$  = pyridine-2-aldoximate;
py = pyridine), characterized  recently (Fig. 1)~\cite{Clerac02}.
At low temperature, this compound can be described
as a chain of ferromagnetic coupled $S = 3$
[Mn$^{III}$-Ni$^{II}$-Mn$^{III}$] units. By comparing
the relaxation time obtained from AC susceptibility
with the one deduced from DC measurements
(relaxation as a function of time), we show that
a unique relaxation time is found over 10 decades.
We demonstrate a quantitative agreement between the theory
and the experiment when the values
of $J$ and $D$ obtained from independent thermodynamical
measurements are considered. Moreover we show that finite-size
effects should be considered to account for the crossover observed on both
the magnetic susceptibility and the relaxation time.

We first present a brief theoretical
description of the expected relaxation
in a SCM. Critical slowing down of the magnetization
is expected in the vicinity of any magnetic critical point.
However, the 1D case is singular as only short-range
order can be observed at finite temperatures.
These correlations affect the dynamic response
of the system and a slowing down of the relaxation
is expected. The 1D Ising model discussed by
Glauber~\cite{Glauber63} gives a single
time process (i.e., a Debye relaxation) with
an exponential enhancement of the relaxation
time that reads (for a spin state $S$ and with
the notations of Eq. 1):
$\tau(T) = \tau_i(T)\exp(8JS^2/k_{\rm B}T)$,
where $\tau_i(T)$ is the individual relaxation
time of each magnetic unit. In this relation,
the exponential factor is directly related to the
temperature dependence of the correlation functions.
At low temperatures, the corresponding energy gap
for the correlation length or the magnetic
susceptibility is $4JS^2$~\cite{note3}.
For any negative value of $D$,
the correlation length remains exponentially
enhanced at low temperatures~\cite{Loveluck75}
and we can still expect a Glauber relaxation.
The value of the corresponding gap
is the energy needed to create a Bloch wall.
It remains the same as in the Ising limit as long
as $|D| \ge 4J/3$~\cite{Barbara94}.
Moreover, each magnetic unit exhibits
uniaxial anisotropy which creates a barrier
as in the SMM case. For magnetic units with
a spin ground state $S$, we therefore expect $\tau_i(T)$
to vary as
$\tau_i(T) = \tau_0\exp(|D|S^2/k_{\rm B}T)$
when quantum tunneling can be ignored. Finally,
we expect the relaxation time to vary as :
\begin{equation}
       \tau(T) = \tau_0\exp((8J + |D|)S^2/k_{\rm B}T)
\label{tau}
\end{equation}
The second point to consider is the influence of defects along the chain.
Finite-size scaling of the Glauber model has been
discussed~\cite{Kamphorst,Luscombe}. In the case of open chains of
size L, a crossover on the relaxation time has been predicted when
the magnetic correlation length of the infinite chain, $\xi$, becomes
of the order of $L$ :
$\tau_{\rm L} (\xi) = \tau_{\infty} (\xi) f_{\tau}(L/\xi)$ (where the
$f_{\tau}$ is the finite-size scaling function introduced in Ref.
~\cite{Kamphorst}). Following this model, the activation energy of 
the relaxation
time should decrease from $(8J + |D|)S^2$ to $(4J + |D|)S^2$ at the
crossover. At approximately the same temperature, we expect a
saturation of $\chi T$~\cite{note3}.

\begin{figure}
\begin{center}
\includegraphics[width=.45\textwidth]{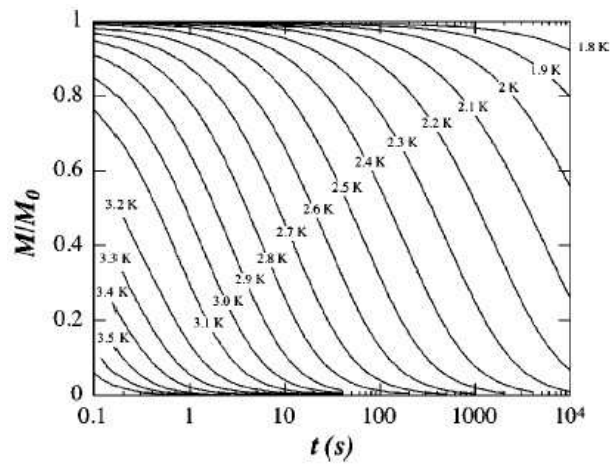}
\caption{Relaxation of the normalized 
magnetization from DC measurements.}
\label{relax}
\end{center}
\end{figure}

The studied SCM is a heterometallic
chain of Mn$^{III}$ and Ni$^{II}$ metal ions (Fig. 1).
The magnetic unit is a [Mn$^{III}$-Ni$^{II}$-Mn$^{III}$]
trimer with strong Ni$^{II}$-Mn$^{III}$
antiferromagnetic interactions
($J_{\rm AF}/k_{\rm B}$ = -21 K)~\cite{Clerac02}.
At low temperature ($k_{\rm B}T \ll J_{\rm AF}$),
the trimers can be described as effective $S = 3$
units connected within the chains by weaker
ferromagnetic interactions (see Fig. 1).
The 3D organization of these chains in the
solid implies that they can be considered
as magnetically isolated.
The magnetic susceptibility of this compound
becomes anisotropic with almost
uniaxial symmetry below 50 K. Between 3.5 K and 7 K,
a frequency dependent AC susceptibility has
been systematically observed in both real and
imaginary components~\cite{Clerac02}.
It approximately corresponds to a single relaxation time
since a satisfactory fit can be obtained
using a generalized Debye model with a small $\alpha$ value ($\approx$ 0.05)
~\cite{Hitoshi03}.
Below 3.5 K, the relaxation becomes too slow
to be observed from AC susceptibility but
DC measurements can still be performed.
With this technique, we have been able to
follow the relaxation down to 1.8 K,
as summarized in Fig. 2. The same shape of the normalized
magnetization $m(t)$ is obtained for all the
investigated temperatures, i.e. the data can be scaled into a single
master curve. As shown in the inset of Fig. 3, a satisfactory fit
can be obtained using a stretched exponential, i.e.
$m(t) = a \exp(-(t/\tau)^{\beta})+b$. The $b$ parameter accounts for a
residual slower relaxation which only concerns a small part of the crystal
($b$ is always less than 5\% at any temperature). $\beta$ is always close
to 0.8 suggesting that some poly-dispersity may be present.
The deduced relaxation time is depicted in Fig. 3.
Note that a consistent value of $\tau$ is obtained simply taking
the time when the normalized magnetization has reached the value $1/e$.
Taken together, AC and DC data show that
a crossover is observed around 2.7 K.
Indeed, above this temperature the dependence
of the relaxation time follows an activated
law with an activation energy of
$\Delta_1/k_{\rm B} \approx 74 K$.
Below 2.7 K, a departure from this simple behavior is observed and a smaller
activation energy; $\Delta_2/k_{\rm B} \approx 55 K$ is found around 2 K
(see Fig. 3).

\begin{figure}
\begin{center}
\includegraphics[width=.45\textwidth]{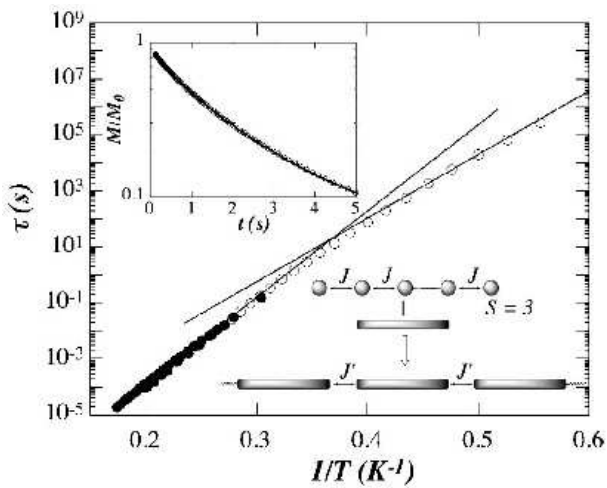}
\caption{Semi-log plot of the relaxation time $\tau$
versus $1/T$. The full and open
dots were obtained from AC and DC measurements,
respectively. The corresponding
straight lines give the energy gaps:
$\Delta_1/k_{\rm B} \approx 74 K$ and $\Delta_2/k_{\rm B} \approx 55
K$. Inset: relaxation of the magnetization at 3 K. Solid line
is the best fit obtained
with the model defined in the text (top). Schematic view of the
magnetic interactions for the model described in the text (bottom).}
\label{fig3}
\end{center}
\end{figure}

\begin{figure}
\begin{center}
\includegraphics[width=.45\textwidth]{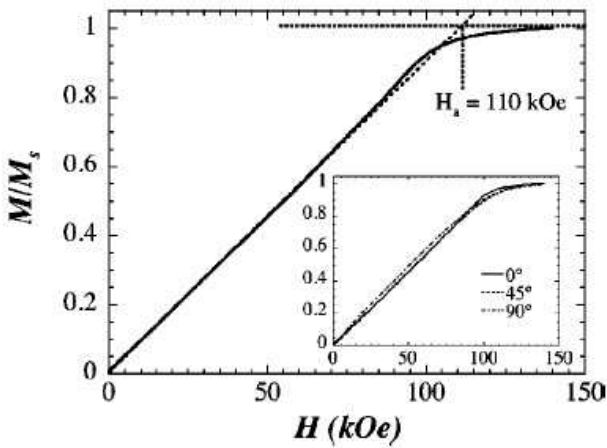}
\caption{Field dependence of the magnetization
(normalized at the saturation value)
when the magnetic field is applied perpendicular
to the easy axis at 1.5 K. The intersection between the
two dotted lines gives the anisotropy field. Inset:
Similar measurements than the main figure
at three angles in the plan normal to the easy axis.
The main figure is at $\theta$ = 0${^\circ}$.}
\label{fig4}
\end{center}
\end{figure}

Using independent magnetic measurements, we have
determined  the anisotropy parameter, $D$,
by measuring the magnetization of a single crystal
as a function of a magnetic field applied perpendicular to the easy axis
(Fig. 4). We obtain a crossover between a linear dependence of
the magnetization at low fields and a saturation at higher
fields. Moreover, the data are almost the same for
any direction perpendicular to the easy axis (inset of Fig.4).
This behavior is characteristic of the uniaxial
anisotropy. An anisotropy field of 110 kOe is
deduced from these data which gives
$D/k_{\rm B}$ = -2.5 K~\cite{note4}. The
temperature dependence of the magnetic susceptibility
is given in Fig. 5 ($\chi T$ versus $1/T$
in a semi-log plot). Full dots were obtained from
AC measurements at low frequency (i.e., below the
characteristic frequency of the relaxation) with a
Quantum Design SQUID magnetometer. Open dots were obtained
from DC measurements with a micro-SQUID experiment~\cite{Wernsdorfer01}.
The expected activated behavior is in fact observed above 6 K (Fig. 5).
The deduced gap, $4JS^2/k_{\rm B} = 28 K$ gives
an estimation of the exchange energy $J/k_{\rm B} = 0.78 K$~\cite{note5,note6}.
This gives $(8J + |D|)S^2/k_{\rm B} = 78 K$ being
in perfect agreement with the
activation energy of the relaxation time, $\Delta$, measured above 2.7 K
(Eq. 2).

Concerning the low temperature behavior, $\chi T$ reaches a
maximum around 5 K, while no anomaly was found on the
relaxation time. This observation precludes
the occurrence of a magnetic phase transition.
On the other hand, this result can be explained
considering finite-size effects (i.e., the presence
of structural defects on the chains which limit the
growth of the correlation length). In fact the
magnetic susceptibility of a finite Ising chain
can be calculated and leads to a saturation
of $\chi T$ at low temperature. To account for the weak decrease of
$\chi T$ observed below 5 K, weak antiferromagnetic interaction,
$J'$, can be introduced between finite chains of $n$ spins (as shown in
inset of Fig. 3)~\cite{note66}. The corresponding susceptibility can
be also determined~\cite{note7} and gives an excellent agreement to
the experimental data below 15 K, where the trimers can be considered
as a $S = 3$ units (Fig. 5). The deduced parameters are the
interaction $J'/k_{\rm B} \approx - 47 mK$ and the average number of
spins between two defects, $n \approx 110$ corresponding to an
average distance, $L \approx 140 nm$.
A more precise discussion can
be made using the field dependence of the magnetization
shown at 4 K in the inset of Fig. 5. The same model
(solid line, inset of Fig. 5) gives a reasonable fit
with $n$ = 90, in good agreement with the value deduced from the temperature
dependence of the susceptibility.
As discussed previously, the finite-size effect should also influence
the relaxation time and the model of~\cite{Luscombe}
leads to a crossover at
a temperature where the $\chi T$ product is maximum. In fact, Fig. 3
shows a crossover on the relaxation time, but at a smaller
temperature, $\approx 2.7 K$, compared to 5 K for the maximum of
$\chi T$. According to the analysis of the susceptibility
measurements, we have generalized the model developed by Luscombe et
al. taking into account the weak $J'$ interactions~\cite{note66}. The
crossover on the relaxation time now occurs at $k_{\rm B}T^{\star} =
(4J - 4J')S^2/\ln (2n)$, i.e. is shifted at lower temperature for an
ferromagnetic interaction, while the activation energy below
$T^{\star}$ becomes $\Delta_2/k_{\rm B} = (4J + 4J' + |D|)S^2$. To
obtain $T^{\star} = 2.7 K$, $J'/k_{\rm B}$ has to be equal to $0.37
K$. This value is not in agreement with $J'/k_{\rm B} = -0.047 K$
obtained from the susceptibility. Nevertheless, the activation energy
$\Delta_2/k_{\rm B} \approx 55 K$ found around 2 K is close to $(4J +
|D|)S^2$ and supports a small $J'$ value.

\begin{figure}
\begin{center}
\includegraphics[width=.45\textwidth]{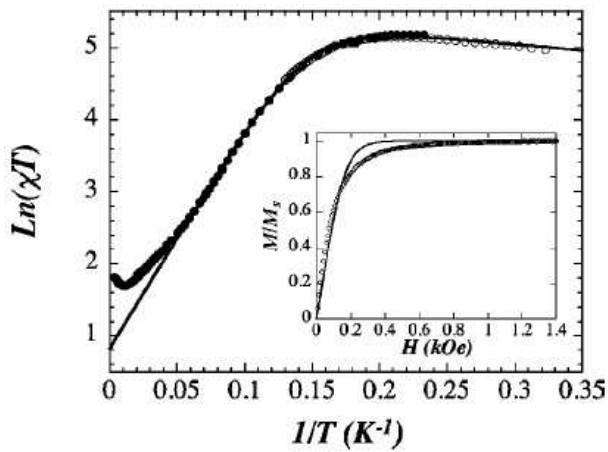}
\caption{Semi-log plot of $\chi T$ versus $1/T$
($\chi$ is given in emu/mol). The full and open
dots were obtained from AC and DC measurements,
respectively. Inset: Field dependence of the magnetization (normalized
at the saturation value) at 4 K and with a field sweep rate of 0.14 kOe/s.
Solid lines correspond to the fit using the expressions given in Ref. [19]}
\label{fig5}
\end{center}
\end{figure}

In summary, the activation energy of a SCM
is a combined effect of the individual kinetics of a trimer
composing the chain and of the magnetic
correlations. The energy parameters
$J$ and $D$ deduced from the relaxation time are
in excellent agreement with the values independently
obtained from magnetic measurements.
A crossover is observed at low temperature
when the magnetic correlation length
becomes of the order of the distance between two defects. A model of these
defects in terms of finite-size chains coupled with weaker exchange
interactions
gives a coherent description of the temperature dependence of the
magnetic susceptibility. It also qualitatively explains the
temperature dependence of the relaxation time although additional
ingredients like poly-dispersity on the chain length or exchange
interaction may be relevant to improve the discussion of the dynamic
properties.

H. M. is grateful for a financial support
from PRESTO project, Japan Science and
Technology Agency (JST). C.C.
and R.C. acknowledge the financial support
of CNRS, Conseil R\'egional d'Aquitaine, and the Universit\'e Bordeaux 1.


\begin{thebibliography}{10}
\bibitem{note60}
Weak interchain couplings and quantum effects should also be mentioned
but will not be considered in the following communication as they remain
negligible in the range of temperature presently explored.
\bibitem{note3}
In any classical problem, $\chi T$ (where $\chi$ is
the magnetic susceptibility) is proportional to the
correlation length $\xi$. Therefore, this result
is true for any value of $D$ as long as the
spins are considered as classical.
\bibitem{note4}
The minimization of the sum of the anisotropy 
and Zeeman energies gives:
$2D S^2 = g\mu_{\rm B}SH_{\rm a}$, where $\mu_{\rm B}$ is the Bohr magneton. 
The $g$ factor is close to 2 in the present case.
\bibitem{note5}
In fact, we are in the Ising limit as our
estimation of $D/J \approx 3$ is larger
than 4/3 (Ref. [12]).
\bibitem{note6}
This estimation is also in agreement with the
mean field analysis of the high temperature susceptibility (Ref. [4]).
\bibitem{note66}
The obtained value of $J'$ may be over estimated as antiferromagnetic
inter-chain coupling may also be present. However this coupling
should remain very weak, i.e. negligible, as no magnetic phase transition
has been observed above 40 mK.
\bibitem{note7}
The expression of the magnetic susceptibility for the model shown in
inset of Fig. 3 is: $\chi = \chi_{\infty} - \chi_{\infty} 
(1 - \exp(-2n \exp(-2K))(1-\tanh K')/
(2n \exp(-2K)(1- \tanh K' \exp(-2n \exp(-2K))))$ where
$\chi_{\infty} = exp(2K)$ is the susceptibility of the infinite
chain, $K=2 JS^2/ k_{\rm B}T$, $K'=2 J'S^2/ k_{\rm B}T$. Below the
maximum of $\chi T$ when $2n \exp(-2K) \ll 1$, this expression
reduces to $\chi = nC \exp(2K')/T$, where $C$ is the Curie constant for
a trimer unit. It corresponds to the susceptibility of a chain of
weakly coupled (by $J'$) saturated segments of n spins. So, in the
same domain of temperatures, the magnetization normalized at its
saturation value reads:
$m_n = \sinh(n g\mu_{\rm B}SH/k_{\rm B}T)/( sinh^2 (n g\mu_{\rm
B}SH/k_{\rm B}T)+exp(-4K')) ^{1/2} $.
This expression has been used to fit the data inset Fig. 5 (solid line).
\end{thebibliography}
\end{document}